\documentclass[prl,a4paper,twocolumn,showpacs ]{revtex4}
\usepackage{amsmath,amsfonts,bbm, graphicx,color}
\def\>{\rangle}
\def\<{\langle}
\def\E{ {\cal E} }

\def\Tr{ \mbox{Tr} }

\begin{document}

\title{Comment on `Quantum resolution to the arrow of time dilemma'} %

\author{David Jennings}%
\affiliation{Institute for Mathematical Sciences, Imperial College London, London SW7 2BW,
United Kingdom}%
\author{Terry Rudolph}%
\affiliation{Institute for Mathematical Sciences, Imperial College London, London SW7 2BW,
United Kingdom}%

\date{September 2009}


\maketitle


In \cite{macc08} it is claimed that entropy decreases can occur, but that any such decrease necessarily coincides with an erasure of memory. It is argued that this resolves the directionality of the arrow of time, since we can only ever have records of entropy increasing events.

Specifically, a \textit{memory of an event} $E$ is defined \cite{macc08} as a physical system $A$ that has a non-zero \emph{classical} mutual information with a system $C$ that bears the consequences of event $E$. For the sake of discussion we adopt this formulation. A purifying environment $R$ may be assumed and it is clear that if the entropy of $A$ and $C$ is to decrease, with no entropy change in the reservoir, $\Delta S(\rho_R)=0$, then the quantum mutual information between $A$ and $C$ must decrease.

Given this setting, the principal claim in \cite{macc08} is that\textit{``any decrease in entropy of a system that is correlated with an observer entails a memory erasure of said observer"}(*).
In a classical setting (*) is true by definition of the entropies - classical events involving the reduction of local entropies trivially coincide with a reduction of memory records, since these are defined as the classical correlations between the memory system $A$ and system $C$.
However, the extension of the argument to all quantum mechanical states, where there is a much richer correlation structure, is non-trivial. Consequently, in this fuller setting the claim is really that entropy decreasing events always coincide with a reduction in classical mutual information \emph{in spite} of the freedoms of quantum mechanics.

This result can only follow if a reduction in the quantum mutual information $I_q(A:C)$ implies a reduction in the classical mutual information $I_c(A:C)$ (the argument in \cite{macc08} proves only that $I_q(A:C)$ is an upper bound on $I_c(A:C)$).
If this were true, since $I_c(A:C)$ is identified with the memory recorded in $A$ of the effects on $C$, it would follow that a reduction in entropy for $C$ can only happen with an erasure of memory of the event for $A$. We note that demanding an entropy-decreasing event $E$ is very different from demanding an event in which all correlations are completely eliminated. For the latter, memory erasure occurs \emph{by assumption} and needs no proof.

An indication that the argument of  \cite{macc08} may be incomplete, is given by the fact that it is possible to have a reduction of $I_q(A:C)$ without a reduction in $I_c(A:C)$. A simple example is sending one particle of a maximally entangled Bell pair through a dephasing channel - in the extreme case of maximal dephasing, $I_q(A:C)$ is reduced to $I_c(A:C)$, but $I_c(A:C)$ certainly remains unchanged. We now argue that instead of resolving the arrow of time dilemma, quantum mechanics actually allows the reduction of local entropies while the classical correlations can \emph{increase}.

The specific example that we consider is a 3 qubit $W$ state, $|\Phi_{ACR} \> = \frac{1}{\sqrt{3}} (|001\> +|010\> +|100\> )$ and the event $E$ that we consider is the action of a CNOT gate controlled on $R$ with $C$ being the target.
Before the event the reduced state $\rho_{AC,i} = \frac{1}{3} |00 \>\< 00 |+\frac{2}{3} | \psi^+ \>\< \psi^+ |$, while after the CNOT event the reduced state is $\rho_{AC,f} = \frac{1}{3}|0 \>\< 0 | \otimes |1 \>\< 1 |+ \frac{2}{3}|+ \>\< + | \otimes |0 \>\< 0 | $. It is straightforward to see that the entropy of the reservoir $R$ or system $A$ does not change, $\Delta S_R=\Delta S_A=0$, however $ \Delta S_C=-0.3683$ and so the quantum mutual information changes from $I_q(A:C ; i) =0.9183 $ to $I_q(A:C;f)=0.5500$.

The calculation of the classical mutual information is a bit involved. Numerics show that for both $\rho_{AC,i}$ and $\rho_{AC,f}$ the optimal measurements are actually projective. It can then readily be derived that for the case of $\rho_{AC,i}$, the optimal measurements are projections in the $| \pm \>$ basis, for which the classical mutual information is $I_c(A:C;i)=0.3499$.
For the final state $\rho_{AC,f}$, the optimal projective measurements for the classical mutual information are straightforward to deduce. The optimal measurement on $C$ is clearly to measure in the $|0\>,|1\>$ basis, on $A$ we simply maximize the discrimination of $|0\>$ and $|+ \>$ with priors of $1/3$ and $2/3$ respectively, which is done by the projective measurement onto $\Pi_{\pm} =(I \pm \cos \theta Z \mp \sin \theta X)/2$ with $\theta = \arctan (\frac{\sqrt{5} - 1}{2})$. One then finds $I_c(A:C;f)=0.3683$, which is higher than the initial classical correlations, despite the reduction of local entropy for $C$.

Instead of quantum mechanics resolving the fact that we have no memory records of entropy decreasing events it actually, in some sense, makes the issue worse.

In some ongoing work we have been investigating the relationship between quantum entanglement and the thermodynamic arrow of time, and in \cite{jennings09} we discuss a `hierarchy of arrows' that arises from the different correlations that can exist in a quantum state.

\newpage
\appendix

\section{Extended Discussion}

We present here an extended version of the single-page comment above.

We present three points on the argument. Firstly, we clarify what is actually being claimed, secondly we show that the argument in \cite{macc08} is incomplete and then thirdly we present a counter-example to the argument.

\subsection{The basic claim}

The general scenario presented in \cite{macc08} involves a memory system $A$, a system $C$ that suffers an event $E$. A purifying environment $R$ is assumed \cite{macc08} and the entropic relation
\begin{eqnarray}\label{entropy}
\Delta S(\rho_A) + \Delta S(\rho_C) - \Delta S(\rho_R) = \Delta I_q(A:C)
\end{eqnarray}
follows immediately from $\Delta S(\rho_R) = \Delta S(\rho_{AC})$ and the definition of the quantum mutual information $I_q(A:C)=S(\rho_A)+S(\rho_C)-S(\rho_{AC})$ for a reduced state $\rho_{AC} = \Tr [\rho_{ACR}]$.
The upshot of this is that if the entropy of $A$ and $C$ is to decrease in the event $E$ (i.e. $\Delta S(\rho_A) + \Delta S(\rho_C)\le 0$) with no entropy exchange with the environment, $\Delta S(\rho_R)=0$, then from (\ref{entropy}) the quantum mutual information $I_q(A:C)$ between $A$ and $C$ must decrease. It is then claimed in \cite{macc08} that memory (in the form of classical mutual information between $A$ and $C$) is always erased in such an event, since the quantum mutual information $I_q(A:C)$ is an upper bound on the classical mutual information $I_c(A:C)$.

The argument in \cite{macc08} says that it applies to \textit{``all physical transformations where entropy is decreased"}, and so for such transformations there is always an associated decrease in the quantum mutual information between the memory and the system $C$. However in a key part of the argument it seems like a very different assumption is made for the event $E$, namely the \textit{``elimination of the quantum mutual information"}. If ``elimination" is meant as the complete removal of quantum mutual information then this is a radically different assumption to an entropy-decreasing event $E$, which only implies a decrease in $I_q(A:C)$.

We must assume that the constraint we impose on the event $E$ is not that all correlations are eliminated, for demanding this would mean that we are demanding an event $E$ such that $\rho_{AC} \rightarrow \rho'_A \otimes \rho'_C$, and for this memory erasure occurs \emph{by assumption} and no proof is required.

The claim of interest is then that any entropy-decreasing event $E$ always coincides with an erasure of some of the classical correlations between $A$ and $C$.

At this point there might be two possible misunderstandings concerning the nature of such a proposed resolution to the arrow of time.
On one hand, a careless reading of the argument might give the impression that the result is `obvious' or `trivially true' and that it lacks any real content. On the other, it is claimed that the resolution is uniquely quantum mechanical.
We argue that in reality both these views are incorrect, and the exact claim amounts to a non-trivial statement about how correlations behave in quantum mechanics.

Classical physics is a special case of quantum physics, and so if the claim were true in general then it would also apply to classical states and classical interactions. In this restricted case the classical states are orthogonal, and the quantum mutual information coincides with the classical mutual information. Consequently, classical events involving the reduction of local entropies will trivially coincide with a reduction of memory records, since these are defined as the classical correlations between the memory system $A$ and system $C$.
Consequently, the argument applies equally well in a classical setting, where it is true almost by definition.

However, the extension of the argument to all quantum mechanical states, where there is a much richer correlation structure, is non-trivial. Consequently, in this fuller setting the claim is really that entropy decreasing events always coincide with a reduction in classical mutual information \emph{in spite} of the freedoms of quantum mechanics.

\subsection{Reduction of $I_q(A:C)$ without the reduction of $I_c(A:C)$}
In this section we show that the argument in \cite{macc08} for (*) is incomplete. The main issue lies in the requirement that reducing the quantum mutual information $I_q(A:C)$ also reduces the classical mutual information $I_c(A:C)$.
The quantum mutual information $I_q(A:C)$, defined as $I_q(A:C) = S(\rho_A) + S(\rho_C) - S(\rho_{AC})$, contains contributions from both classical correlations as well as purely quantum mechanical correlations. We may reduce the purely quantum correlations without affecting the classical ones, and so the claim (*) is called into question.

For example, consider the pure, maximally entangled bipartite state $\rho_{AC}= |\Phi^+\>\<\Phi^+|$.
\begin{eqnarray}
|\Phi^+\> &=& \frac{1}{\sqrt{2}} (|01 \> + |10\> ).
\end{eqnarray}

Initially, $I_q(A:C) =2$. The classical mutual information between $A$ and $C$ is defined as $I_c= \mbox{max}_{E_i \otimes F_j} H(E_i:F_j)$, where
\begin{eqnarray}\label{classical}
H(E_i:F_j)&=& \sum_{ij} P_{ij} \log P_{ij}-\sum_i p_i \log p_i\nonumber\\
 &&- \sum_j q_j \log q_j
\end{eqnarray}
and $P_{ij} =\Tr [E_i \otimes F_j \rho_{AC}]$, $p_i=\sum_j P_{ij}$ and $q_j=\sum_i P_{ij}$ given POVMs $E_i$ and $F_j$ for $A$ and $C$ respectively. For projective measurements in the computational basis we find that $I_c(A:C) =1$.

If we apply the dephasing channel
\begin{eqnarray}
\E (\rho_{AB}) &=& (1-p) \rho_{AB} + p Z_A\rho_{AB}Z_A
\end{eqnarray}
then it is easy to determine that $I_q(A:C) = 2- h(p)$, with $h(p) = -p \log p - (1-p)\log (1-p)$, however $I_c(A:C)=1$ since perfect classical correlations still remain for local projective measurements in the computational basis.
In the case of $p=1/2$, the state is a product state and only the classical correlations remain.
Consequently, it is possible to reduce $I_q(A:C)$ without reducing $I_c(A:C)$.

\subsection{A decrease in local entropy with an increase in classical correlations}
The example in the previous section suggests that the main argument (*) of \cite{macc08} is incomplete. We now argue that (*) is incorrect and that instead of resolving the arrow of time dilemma, quantum mechanics actually allows the reduction of local entropies while the classical correlations can \emph{increase}.  The intuitive idea is that reducing the quantum correlations between $A$ and $C$ corresponds to the local reduction of entropy, with some of the quantum correlations converted into classical correlations in the process.

The specific example that we consider is a 3 qubit $W$ state, $|\Phi_{ACR} \> = \frac{1}{\sqrt{3}} (|001\> +|010\> +|100\> )$ and the event $E$ that we consider is the action of a CNOT gate controlled on $R$ with $C$ being the target.

Before the event the reduced state $\rho_{AC,i} = \frac{1}{3} |00 \>\< 00 |+\frac{2}{3} | \psi^+ \>\< \psi^+ |$, while after the CNOT event the reduced state is $\rho_{AC,f} = \frac{1}{3}|0 \>\< 0 | \otimes |1 \>\< 1 |+ \frac{2}{3}|+ \>\< + | \otimes |0 \>\< 0 | $.

It is straightforward to see that the entropy of the resevoir $R$ or system $A$ does not change, $\Delta S_R=\Delta S_A=0$, however $ \Delta S_C=h(1/2+\sqrt{5}/6)-h(2/3)=-0.3683$ and so from (\ref{entropy}) we have that the quantum mutual information changes from $I_q(A:C ; i) =h(2/3)=0.9183 $ to $I_q(A:C;f)=h(1/2+\sqrt{5}/6)=0.5500$.

The calculation of the classical mutual information is more involved. $I_c(A:C)$ is the maximum over local POVM measurements $E_i \otimes F_j$ of the Shannon mutual information for the distributions $P_{ij} = \Tr [E_i \otimes F_j \rho_{AC}]$.

This classical mutual information is the largest accessible information for $C$ given the ensemble $\{q_j , \rho^j_C\}$, prepared by $A$ with the POVMs $\{E_i\}$. Equivalently, it is the largest accessible information for $A$ given the ensemble $\{p_i , \rho^i_A\}$ prepared by $C$ with the POVMs $\{F_j \}$. Furthermore, the extraction of the accessible information can be achieved using rank 1 POVMs\cite{davies78}.

For the initial state $\rho_{AC,i}$ we must consider the joint probability distribution
\begin{eqnarray}
p_{ij}&=& 1/3 (|\< \tilde{\psi}_i | 0 \>|^2|\< \tilde{\phi}_j | 0 \>|^2+|\< \tilde{\phi}_j |\sigma_x | \tilde{\psi}_i \>|^2
\end{eqnarray}
with the POVMs for $A$ and $C$ being respectively $E_i = |\tilde{\psi}_i \>\< \tilde{\psi}_i |$ and $F_i = |\tilde{\phi}_j \>\< \tilde{\phi}_j |$ with $|\tilde{\psi}_i \>$ and $|\tilde{\psi}_i \>$ subnormalized.

The optimal projective measurements for $A$ and $C$ are to measure in the $| \pm \>$ basis, for which the classical mutual information is $I_c(A:C;i)=2-\log 3 - \log 4 + (5/6) \log 5 = 0.349$ and furthermore numerics show that non-projective POVMs do not yield a higher mutual information.

For the final state $\rho_{AC,f}$, the optimal local measurements for the classical mutual information are straightforward to deduce. They are again projective measurements and take the form
\begin{eqnarray}
E_1 &=& (1/2)(I + \cos \theta Z - \sin \theta X) \nonumber \\
E_2 &=& (1/2)(I - \cos \theta Z + \sin \theta X) \nonumber \\
F_1 &=& (1/2)(I + Z) \nonumber \\
F_2 &=& (1/2)(I - Z)
\end{eqnarray}
where $\theta = \arctan (\frac{\sqrt{5} - 1}{2})$. These are the optimal measurements since $C$ must measure on their orthogonal, and hence distinguishable, states and then $A$ maximizes the discrimination of $|0\>$ and $|+ \>$ with priors of $1/3$ and $2/3$ respectively. Consequently, 
\begin{eqnarray}
P_{11}=(2/3)(1/2 +1/\sqrt{5})&,&P_{12}=(2/3)(1/2 -1/\sqrt{5}) \nonumber \\ 
P_{21}=(1/3)(1/2 +1/2\sqrt{5})&,&P_{22}=(1/3)(1/2 -1/2\sqrt{5})\nonumber
\end{eqnarray} 
and so from (\ref{classical}) we have $I_c(A:C;f)=0.368$, which is higher than the initial classical correlations, despite the reduction of local entropy for $C$.

\subsection{Conclusions}
We have investigated the claims of \cite{macc08} and argued that instead of being a trivial result or one that is uniquely quantum mechanical, they are claims about how classical correlations as a subset of quantum correlations behave under reductions of entropy.

We demonstrated, through the example of a dephasing channel on the $|\psi^+ \>$ state, that one may reduce purely quantum mechanical correlations while keeping the classical correlations intact. Furthermore, we gave an example in which a composite system $| W \>_{ACR}$ undergoes a unitary transformation for which $\Delta S_R =\Delta S_A= 0$, $\Delta S_C = -0.3683$ while the classical mutual information, $I_c(A:C)$, defined as an optimization over local measurements at the memory system $A$ and the system $C$, increases from $0.349$ to $0.368$. In other words we have that the local entropy of $C$ decreases, while the classical correlations between the memory and the system that suffers the event $E$ actually increase rather than decrease.
Nothing forbids part of the initial, purely quantum mechanical, correlations between $A$ and $C$ being used up to produce a decrease in local entropy at $C$ and at the same time produce an increase in the classical correlations between $A$ and $C$.
Of course any event in which all of the correlations are removed would clearly remove any classical correlations present, but these extreme cases are a special subset of all entropy-decreasing events where memory erasure is true by assumption.
Consequently, instead of quantum mechanics resolving the fact that we have no memory records of entropy decreasing events it actually, in some sense, makes the issue worse.

In some ongoing work we have been investigating the relationship between quantum entanglement and the thermodynamic arrow of time, and in \cite{jennings09} we discuss a `hierarchy of arrows' that arises from the different correlations that can exist in a quantum state.
\subsection{Acknowledgements}
We thank Sean Barrett for useful comments on an earlier draft.

\end{document}